\documentclass{article}

\usepackage{arxiv}
\PassOptionsToPackage{table}{xcolor}  

\usepackage[utf8]{inputenc} 
\usepackage[T1]{fontenc}    
\usepackage{hyperref}       
\usepackage{url}            
\usepackage{booktabs}       
\usepackage{amsfonts}       
\usepackage{nicefrac}       
\usepackage{microtype}      
\usepackage{lipsum}
\usepackage{graphicx}
\usepackage{multirow}
\usepackage{soul}
\usepackage{amssymb}
\usepackage{listings}
\usepackage{amsmath} 
\usepackage{makecell}
\usepackage{subcaption}
\usepackage{tikz}
\usepackage{siunitx}

\usepackage{xcolor} 
\definecolor{azure}{rgb}{0.0, 0.5, 1.0}
\definecolor{darkgreen}{rgb}{0.0, 0.5, 0.0}
\definecolor{amaranth}{rgb}{0.9, 0.17, 0.31}
\definecolor{cadetgrey}{rgb}{0.57, 0.64, 0.69}
\definecolor{aureolin}{rgb}{0.99, 0.93, 0.0}
\graphicspath{ {./images/} }

\usepackage{pgfplots}

\title{LLM \& HPC:Benchmarking DeepSeek's Performance in High-Performance Computing Tasks}

\author{\href{https://orcid.org/0009-0000-4687-1416}{\includegraphics[scale=0.06]{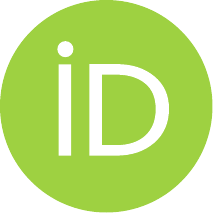}\hspace{1mm}
  Noujoud Nader},  \\
  Center of Computation and Technology\\
  Louisiana State University\\
  Baton Rouge, LA, 70803 \\
  \texttt{nnader@lsu.edu} \\
   \And
 \href{https://orcid.org/0000-0003-3922-8419}{\includegraphics[scale=0.06]{orcid.pdf}\hspace{1mm}Patrick Diehl} \\
  Los Alamos National Laboratory\\
  Los Alamos, NM, 87544 \\
  \texttt{diehlpk@lanl.gov} \\
  \And
 \href{https://orcid.org/0000-0002-7979-2906}{\includegraphics[scale=0.06]{orcid.pdf}\hspace{1mm}Steve Brandt} \\
  Center of Computation and Technology\\
  Louisiana State University\\
  Baton Rouge, LA, 70803 \\
  \texttt{sbrandt@lsu.edu} \\
   \And
 \href{https://orcid.org/0000-0002-8712-2806}{\includegraphics[scale=0.06]{orcid.pdf}\hspace{1mm}Hartmut Kaiser} \\
  Center of Computation and Technology\\
  Louisiana State University\\
  Baton Rouge, LA, 70803 \\
  \texttt{hkaiser@lsu.edu} \\
}

\newcommand{\cpp}[0]{C\texttt{++}}

\begin{document}
%
%
%

%
%

\maketitle              
\begin{abstract}
Large Language Models (LLMs), such as GPT-4 and DeepSeek, have been applied to a wide range of domains in software engineering. However, their potential in the context of High-Performance Computing (HPC) much remains to be explored. This paper evaluates how well DeepSeek, a recent LLM, performs in generating a set of HPC benchmark codes: a conjugate gradient solver, the parallel heat equation, parallel matrix multiplication, DGEMM, and the STREAM triad operation. We analyze DeepSeek's code generation capabilities for traditional HPC languages like \cpp, Fortran, Julia and Python. The evaluation includes testing for code correctness, performance, and scaling across different configurations and matrix sizes. We also provide a detailed comparison between DeepSeek and another widely used tool: GPT-4. Our results demonstrate that while DeepSeek generates functional code for HPC tasks, it lags behind GPT-4, in terms of scalability and execution efficiency of the generated code.

\end{abstract}
\section{Introduction}
Large Language Models (LLMs) have rapidly advanced, potentially approaching Artificial General Intelligence (AGI). Trained on vast textual data, LLMs have shown remarkable capabilities in natural language processing and visualization tasks~\cite{anthropicClaudeSonnet,google,OpenAi}. This trend has accelerated with the release of the DeepSeek models by the Chinese company DeepSeek~\cite{DeepSeek}. Their DeepSeek-V3~\cite{liu2024deepseek,guo2024deepseeka} was trained in two months at a cost of \$5.6 million, significantly cheaper than similar models. In contrast, the development cost of ChatGPT-4 ranged between \$41 million and \$78 million \cite{forbesExtremeCost}. On January 20th 2025, they launched DeepSeek-R1~\cite{guo2025deepseek}, a reasoning model exhibiting performance comparable to OpenAI's\ O1 model, available for open research~\cite{gibney2025china}. DeepSeek-R1 is trained through large-scale reinforcement learning (RL) without supervised fine-tuning, demonstrating powerful reasoning behaviors~\cite{guo2025deepseek}.

In the high-performance computing (HPC) domain, LLMs are being explored for tasks such as code analysis, generation, and optimization. However, there is a lack of standardized, reproducible evaluation processes for LLMs in HPC-specific tasks.
For this reason, this paper explores the performance of the open-source LLM DeepSeek for simple code generation tasks in HPC. We evaluate code generation using four programming languages (PLs) selected for their prominence in HPC and based on the \textit{TIOBE\ index}~\cite{tiobeTIOBEIndex}. The PLs are \cpp\ for performance and simulation, Fortran for numerical and array optimizations, Python for flexibility and integration with other HPC languages, and Julia for scientific computing performance with a Python-like syntax.

We present five examples to assess the performance of the model. The first example in this study features a conjugate gradient solver that incorporates matrix and vector operations. The second example is a parallel 1D stencil-based heat equation solver~\cite{diehl2023benchmarking}, chosen for its relevance in testing parallel computational workflows and its widespread use in scientific computing.
The third example is parallel matrix multiplication, a fundamental operation in scientific computing and machine learning, often accelerated by using HPC systems to enhance computational speed. The fourth example is parallel double-precision general matrix multiply (DGEMM), a widely used operation in numerical linear algebra that is essential for many scientific and engineering applications, testing the system's computational efficiency for large-scale matrix operations. The final example is the STREAM triad operation, a key benchmark for testing memory bandwidth and data movement between CPU and memory, commonly used to evaluate the memory subsystem's performance in HPC systems.
We summarize the main contributions of this paper as follows:
\begin{itemize}
    \item A comparison of DeepSeek and GPT-4 for HPC code generation, specifically targeting five simple tasks.
    \item A systematic evaluation of the performance and quality of code generated by DeepSeek in when writing code for four widely used HPC programming languages: \cpp, Fortran, Python, and Julia.
    \item A comprehensive analysis of the models' ability to generate optimized code for common HPC operations, offering insights into the potential of LLMs for high-performance computing tasks.
\end{itemize}

\begin{table}
\centering
\rowcolors{2}{gray!25}{white}
\caption{Summary of LLM-based HPC related work}
\label{tab:related_work}
\begin{tabular}{c|c|c}
\textbf{Paper} & \textbf{Main Focus} & \textbf{LLM Model Used} \\
\toprule
LLM4HPC~\cite{chen2023lm4hpc} & HPC Adaptation & LLaMa-2 \\
\midrule
LLM4VV~\cite{munley2024llm4vv} & OpenACC Testing & GPT-4, CodeLlama \\
\midrule
HPC-GPT~\cite{ding2023hpc} & \makecell{AI Model \\Management} & LLaMa-13B \\
\midrule
Tokompiler LLM~\cite{kadosh2023scope} & Code Completion & GPT-3 \\
\midrule
HPC-Coder~\cite{nichols2024hpc} & \makecell{OpenMP \\ \& MPI Handling} & DeepSpeed \\
\midrule
\makecell{Dataset for \\ OpenMP Translation~\cite{lei2023creating} }& Code Translation & LLaMa-2 \\
\midrule
LLMs in HPC~\cite{chen2024landscape} & LLM-HPC Challenges & Various LLMs \\
\midrule
Godoy et al. (2023)~\cite{godoy2023evaluation} & Kernel Code Generation & LLaMa-2 \\
\midrule
Valero-Lara et al. (2023)~\cite{valero2023comparing} & LLaMa-2 Comparison & LLaMa-2 \\
\midrule
Godoy et al. (2024)~\cite{godoy2024large} & Code Parallelization & GPT-3 \\
\midrule
chatHPC~\cite{yin2025chathpc} & HPC Chatbot & \makecell{GPT-like \\(based on StarCoder)} \\
\bottomrule
\end{tabular}
\end{table}

\section{Related Work}
LLMs have been widely used across various domains, including natural language processing (NLP) and visualization.  Multiple efforts have also been made to benchmark LLMs for tasks such as code generation \cite{diehl2024evaluating,diehl2025llm,chen2021evaluating}. However, their application in analyzing and optimizing HPC tasks remains challenging. In Table~\ref{tab:related_work}, we present the related works that use LLMs for HPC tasks. LLM4HPC~\cite{chen2023lm4hpc} represents the first effort to adapt an LLM specifically to the HPC domain. The LLM4HPC framework is specifically designed for HPC, showing success in code similarity analysis, parallelism detection, and OpenMP question-answering tasks. LLM4VV~\cite{munley2024llm4vv} is a fine-tuned model that uses the capabilities of GPT-4 and CodeLlama (based on Llama-2) to successfully generate OpenACC directives. HPC-GPT introduced by Ding et al.~\cite{ding2023hpc}, was built on LLaMa-13B. HPC-GPT has been successfully applied to managing AI models and datasets, as well as detecting data races. Kadosh et al.~\cite{kadosh2023scope} introduced the domain-specific Tokompiler LLM, which outperforms a GPT-3-based model in code completion and semantics for Fortran, C, and \cpp\ code. Nichols et al.~\cite{nichols2024hpc} present HPC-Coder. Their work demonstrated varying success in code completion, including handling OpenMP pragmas and MPI calls. Lei et al.~\cite{lei2023creating} introduced a dataset designed for fine-tuning models on OpenMP Fortran and \cpp\ code translation. Their fine-tuned model yielded more accurate results compared to GPT-4. Chen et al.~\cite{chen2024landscape} provide an insightful overview of the challenges and opportunities at the intersection of LLMs and HPC, extending beyond code generation. Notably, the works by Godoy et al.~\cite{godoy2023evaluation} and Valero-Lara et al.~\cite{valero2023comparing}, which evaluate HPC kernel code generation and results for LLaMa-2, are among the first to apply LLM-based code generation to the domain of HPC software development. Additionally, Godoy et al.~\cite{godoy2024large} apply LLM capabilities of GPT-3 targeting HPC kernels for code generation, and auto-parallelization of serial code in \cpp, Fortran, Python and Julia. Yin et al.~\cite{yin2025chathpc} proposed chatHPC, a chatbot for HPC question answering and script generation.

\section{Methodology}
We used five examples: \textit{(1)} a conjugate gradient solver, \textit{(2)} a parallel one-dimensional heat equation solver using finite differencing, \textit{(3)} parallel matrix multiplication, \textit{(4)} parallel DGEMM, and \textit{(5)} the STREAM triad operation. The complexity of the code increases with each example. We generated the code for these examples using DeepSeek-R1 and compared it with our previous results~\cite{diehl2024evaluating} obtained using GPT 4.0 for the first two examples. The code was generated on 02/03/2025. The queries used for code generation are shown in Table~\ref{tab:queries}. 

\begin{table}[ht]
\caption{Queries for HPC tasks code generation, whether the language was C\texttt{++}, Fortran, Julia, Python}
\label{tab:queries}
\rowcolors{2}{gray!25}{white}
\centering
\begin{tabular}{|c|p{2.2cm}|p{9cm}|}
\hline
\textbf{Ex.}& \textbf{Problem}& \textbf{Prompt} \\
\hline
1&Conjugate Gradient Solver & Write a \textbf{language} code to solve the linear equation system using the conjugate gradient solver and validate it.\\
\hline
2& Parallel 1D Stencil-Based Heat Equation Solver&Write a parallel \textbf{language} code to solve the one-dimensional heat equation using a finite difference scheme for the discretization in space and the Euler method
for time integration, validate it and plot the solution.\\
\hline
3&Parallel Matrix Multiplication & Write a parallel \textbf{language} code for matrix multiplication and validate it.\\\hline
4& Double-Precision General Matrix Multiplication& Write a parallel \textbf{language} code to perform DGEMM on large matrices, optimize the implementation for performance using parallel computing techniques, validate the results, and compare the performance with different matrix sizes and parallelization strategies.\\\hline
5& STREAM & Write a parallel \textbf{language} code to perform the STREAM triad operation on large arrays.\\
\hline
\end{tabular}

\end{table}

\textbf{Example 1} : An advanced example from numerical methods textbooks is using a conjugate gradient solver to solve a system of linear equations~\cite{shewchuk1994introduction}.
\begin{align}
    A^{n \times n} \cdot x^n = b^n \quad \text{with} \quad n \in \mathbb{Z}^+, A = A^t, \text{ and } x^t A x > 0, \forall X \in \mathbb{R}^n\text{.} 
\end{align}
To evaluate the code, we asked DeepSeek to use the following equation system $M \cdot x = b$ with 
\begin{align}
M = \left( \begin{matrix}
    4 &  -1 & 0 \\
    -1 &  4 &  -1 \\
    0 &  -1 &  4
\end{matrix}   \right), b= \left( \begin{matrix}
    1.0 \\ 2.0 \\ 3.0
\end{matrix} \right) 
\end{align}
And the correct result is  $x_\text{exact} = \left( \begin{matrix}
    13.0/28.0 \\ 6.0/7.0 \\ 27.0/28.0
\end{matrix} \right) $.

\noindent\textbf{Example 2 : }Here, we want to evaluate whether DeepSeek can write the code that can solve: 
    \begin{align}
       \frac{\partial u}{\partial t} = \alpha\frac{\partial^2 u}{\partial x^2}, \quad 0 \leq x < L, t>0
    \end{align}
    where $\alpha$ is the material's diffusivity. For the discretization in space, a finite difference scheme was used
    \begin{align}
        u(t+1,x_i) = u(t,x_i) + dt \ \alpha \frac{u(t,x_{i-1}) - 2 u(t,x_i) + u(t,x_{i+1})}{h^2}
    \end{align}
    We did not specify how to generate the grid, \emph{i.e.}\ equidistant nodal spacing with $n$ grid points $x = \{ x_i = i \cdot h \in \mathbb{R} \vert i = 0,\ldots,n-1\} $, nor what time integration method to use, \emph{e.g.}\ the Euler method.

\noindent\textbf{Example 3:} In this example, we evaluate DeepSeek's ability to generate parallel code for \textbf{matrix multiplication}, a general benchmarking task used to evaluate the basic performance and scalability of parallel code. It is also a fundamental operation in numerical simulations and machine learning. The task involves writing efficient parallel code that runs across multiple threads or processors, optimizing for memory usage and computational speed. The generated code is supposed to compute the following:
\begin{align}
    C(i,j) = C(i,j) + A(i,k) * B(k,j)\text{,}\quad A,B,C \in \mathbb{R}^{n\times n} \text{.}
\end{align}

\noindent\textbf{Example 4:} The fourth example is more general and focuses on \textbf{DGEMM}, which targets a more specialized application within the HPC domain. The objective here is to analyze the ability of DeepSeek to generate parallel code for performing DGEMM on large-scale matrices, with a particular focus on optimizing for performance using parallel computing techniques. The generated code is supposed to compute the following:
\begin{align}
     C = \alpha\cdot A \cdot B + \beta \cdot C, \quad A,B,C \in \mathbb{R}^{n\times n}, \alpha,\beta\in \mathbf{R} \text{.}
\end{align}
While Example 3 is useful for assessing general parallel performance, DGEMM provides a deeper evaluation of optimized performance for large-scale, high-precision operations.\\
\noindent\textbf{Example 5:} The final example evaluates DeepSeek's ability to implement the \textbf{STREAM} triad operation, which is commonly used in HPC benchmarks to measure memory bandwidth and data throughput. The task requires writing parallel code that performs the STREAM triad operation:
\begin{align}
    A[i] = B[i] + scalar \cdot C[i], \quad A,B,C\in \mathbb{R}^n, scalar\in\mathbb{R}\text{.}
\end{align}
The performance of the implementation is evaluated based on its ability to efficiently utilize memory bandwidth, especially in multi-core or distributed HPC systems. The goal is to assess how well the code is generated, how well it scales as the array sizes grow, and how effectively it can handle large data sets while maintaining high computational throughput.
 
 We copied the generated code to the paper's GitHub repository\footnote{\url{https://github.com/NoujoudNader/AiCode\_DeepSeek}}. For some of the generated codes, DeepSeek provided instructions on how to compile the code and some examples of expected output. These instructions were added to the Github repository.

\section{Quality of the generated code}
All of the generated codes were checked for compilation errors, runtime errors, and correctness. Table~\ref{tab:evaluation:overview} summarizes the evaluation of all examples. For the conjugate gradient and parallel heat equation solver, the generated code with the Deep Seek model is compared with the generated code with ChatGPT 4.0 from the author's previous work~\cite{diehl2024evaluating}.

\begin{table}[tb]
\caption{Results for the generated code performance across two examples (Conjugate Gradient and Parallel Heat Equation). We verify that the code compiled successfully for \cpp and FORTRAN. We verified that the codes executed without any runtime errors, and that the code produced correct results for all languages. A comparison was made between DeepSeek and our previous results with GPT~\cite{diehl2024evaluating}.}
    \centering
    
    \begin{tabular}{c|cccc|cccc}\toprule
       Model  &  \multicolumn{4}{c|}{Deep Seek}  &  \multicolumn{4}{c}{ChatGPT~\cite{diehl2024evaluating}}\\\midrule
        Language &  \cpp & Fortran & Python & Julia &  \cpp & Fortran & Python & Julia\\\midrule
         & \multicolumn{8}{c}{Conjugate gradient}\\\midrule
    Compile     & \checkmark  & x  & -- & -- & \checkmark & \checkmark & -- & -- \\ 
    \rowcolor{lightgray} Execution &  \checkmark & \checkmark & \checkmark & \checkmark  & \checkmark & \checkmark & \checkmark & x\\
    Correctness & \checkmark & \checkmark & \checkmark & \checkmark & \checkmark & \checkmark & \checkmark & \checkmark\\\midrule
    & \multicolumn{8}{c}{Parallel heat equation}\\\midrule
    Compile   & \checkmark & x & -- & -- & x & \checkmark & -- & --\\ 
     \rowcolor{lightgray} Execution & \checkmark & \checkmark & \checkmark & x & \checkmark & \checkmark & x & x \\
    Correctness & \checkmark & \checkmark & \checkmark & \checkmark &  \checkmark & x & \checkmark & \checkmark\\\midrule
    \end{tabular}
    \label{tab:evaluation:overview}
\end{table}

\subsection{Conjugate gradient}
For Python, the generated code executed and produced the correct results. The \cpp\ code compiled, executed, and produced the correct result.  The Fortran code did not compile due to the following \texttt{ Error: Unexpected data declaration statement at (1)} since the exact result was declared in the third last line. Moving the declaration to the top fixed the issue. After that the code compiled, executed, and produced the correct results. 
In Python, the exact solution was computed using \lstinline[language=python]{np.linalg.solve} from the NumPy package, and in the Julia code used \lstinline[language=python]{x_exact = A \ b}. In the other codes the exact solution was hard-coded. The Deep Seek model had one compilation issue in the Fortran code and GPT had no compilation issues. The Julia code had runtime issues for GPT but not for Deep Seek. To summarize, for the conjugate gradient method, each model had one issue with the generated code.

\subsection{Parallel heat equation solver}
The \cpp\ and Python code worked without issue. The Fortran code did not compile since the variable \lstinline[language=fortran]{pi} was used but not declared. After declaring the variable the code worked. The Julia code had the following error \texttt{ERROR: LoadError: UndefVarError: `@printf` not defined in `Main`}. After adding \texttt{using Printf} the code worked. The \cpp\ and Fortran code used OpenMP for parallelism. The Python code used \textit{numba} and Julia used \textit{Base.Threads} for parallelism. The Fortran code did not compile for DeepSeek but did compile for ChatGPT. The Fortran code produced the correct result for Deep Seek but not for ChatGPT. For both models, the Julia code had runtime errors. The Python code had runtime errors for ChatGPT but no errors for DeepSeek. The model added the following parameters: heat coefficient $\alpha =0.1$, length of the bar $L=1$, nodal spacing $h=0.1$, final time $T=1$, and time step width $dt=10^{-3}$.

\subsection{Parallel matrix multiplication}
The matrices were filled with random numbers and the serial and parallel computation were compared. The \cpp, Fortran, and Python codes worked and had correct results with respect to the serial execution. The Julia code had one error while printing the results. The \cpp\ and Fortran code used OpenMP, the Python code used the \textit{multiprocessing} package, and the Julia code used \textit{Base.Threads} package.

\subsection{DGEMM}
The Fortran code did not compile due to \texttt{passed REAL(4) to REAL(8)} and the result for the parallel implementation was incorrect. The Python code stopped execution with \lstinline[language=python]{numba.core.errors.UnsupportedRewriteError}. After investigating the error, we discovered that the model generated the code using \texttt{nb.prange(0, n, block\_size)}, however, the Python package \textit{numba} does not provide these arguments. To fix the code, we needed to edit the \textit{numba} API calls. The \cpp\ and Fortran code used OpenMP and Julia used the package \textit{Base.Threads}.

\subsection{STREAM}
The \cpp\ code did not compile due to SIMD errors \texttt{error: `c' in `aligned' clause is neither a pointer nor an array nor a reference to pointer or array}. The Fortran code did not compile due to \texttt{Error: Unclassifiable statement at (1)}. The Python code stopped with the error \texttt{AttributeError: 'c\_double\_Array\_100000000' object has no attribute 'get\_obj'}. The Julia code had the following error \texttt{ERROR: LoadError: UndefVarError: `nthreads` not defined in local scope}. The \cpp\ code used OpenMP, the Fortran code used coarray, the Python code used the \textit{multiprocessing} package, and Julia used the \textit{Base.Threads} package.

\begin{table}[tb]
 \caption{Results for the performance of the generated code across three HPC examples (Matrix Multiplication, DGEMM, and Stream). We verified that the code compiled successfully for \cpp\ and FORTRAN. We verified that the codes executed without any runtime errors, and that the code produced correct results for all codes.}
    \centering
    \rowcolors{2}{gray!25}{white}
    \begin{tabular}{l|cccc|cccc|cccc}
      Language & \rotatebox{60}{\cpp} & \rotatebox{60}{Fortran} & \rotatebox{60}{Python} & \rotatebox{60}{Julia} &
      \rotatebox{60}
      {\cpp} & \rotatebox{60}{Fortran} & \rotatebox{60}{Python} & \rotatebox{60}{Julia} & 
      \rotatebox{60}{\cpp} &\rotatebox{60}{Fortran} & \rotatebox{60}{Python} & \rotatebox{60}{Julia}\\\midrule
     Example &  \multicolumn{4}{c|}{Matrix Multiplication}   &  \multicolumn{4}{c|}{DGEMM} &  \multicolumn{4}{c}{Stream}  \\\midrule
     Compilation & \checkmark & \checkmark & -- & -- & \checkmark & x & -- & -- & x & x & -- & --\\
     Execution & \checkmark & \checkmark & \checkmark & x & \checkmark & \checkmark & x & x & \checkmark & \checkmark & x & x\\
     Correctness & \checkmark & \checkmark & \checkmark & \checkmark & \checkmark & x & ? & \checkmark & \checkmark & \checkmark\ & \checkmark & \checkmark\\\bottomrule
    \end{tabular}
    \label{tab:quality:2}
\end{table}

\begin{figure}[tb]
    \centering
    \begin{subfigure}[t]{0.28\textwidth}
    \begin{tikzpicture}[scale=0.7, transform shape]
    \draw[help lines, color=gray!30, dashed] (-0.1,-0.1) grid (2.9,2.9);
    \draw[->,thick,cadetgrey] (0,0)--(3,0) node[right]{Difficult};
    \draw[->,thick,cadetgrey] (0,0)--(0,3.1) node[above,cadetgrey]{Good};
    \node[left,cadetgrey] at (0,0) {Easy};
    \node[below,cadetgrey] at (0,0) {Poor};
    \draw[->,thick,cadetgrey] (0,0)--(3,0) node[right]{Difficult};
    \draw[->,thick,cadetgrey] (0,0)--(0,3.1) node[above,cadetgrey]{Good};
    \draw[fill=gray] (2.069767442,3) circle [radius=0.05] node[right] {py}; 
    \draw[fill=gray] (3,3) circle [radius=0.05] node[above] {\cpp};
    \draw[fill=gray] (2.302325581,2) circle [radius=0.05] node[left] {Fortran}; 
    \draw[fill=gray] (2.069767442,3) circle [radius=0.05] node[above] {jl};
    \draw[fill=azure] (2.069767442,3) circle [radius=0.05] node[below] {\textcolor{azure}{py}}; 
    \draw[fill=azure] (1.583850932,3) circle [radius=0.05] node[below] {\textcolor{azure}{\cpp}};
    \draw[fill=azure] (2.637931034,2) circle [radius=0.05] node[right] {\textcolor{azure}{Fortran}}; 
    \draw[fill=azure] (2.405172414,1.5) circle [radius=0.05] node[above] {\textcolor{azure}{jl}};
    \end{tikzpicture}
    \caption{Conjugate gradient}
    \label{fig:cocomo:cg}
    \end{subfigure}
    \hfill
    \begin{subfigure}[t]{0.28\textwidth}
    \begin{tikzpicture}[scale=0.7, transform shape]
    \draw[help lines, color=gray!30, dashed] (-0.1,-0.1) grid (2.9,2.9);
    \draw[->,thick,cadetgrey] (0,0)--(3,0) node[right]{Difficult};
    \draw[->,thick,cadetgrey] (0,0)--(0,3.1) node[above,cadetgrey]{Good};
    \node[left,cadetgrey] at (0,0) {Easy};
    \node[below,cadetgrey] at (0,0) {Poor};
    \draw[->,thick,cadetgrey] (0,0)--(3,0) node[right]{Difficult};
    \draw[->,thick,cadetgrey] (0,0)--(0,3.1) node[above,cadetgrey]{Good};
    \draw[fill=gray] (2.752293578,3) circle [radius=0.05] node[below] {py}; 
    \draw[fill=gray] (3,3) circle [radius=0.05] node[above] {\cpp};
    \draw[fill=gray] (2.779816514,2) circle [radius=0.05] node[right] {Fortran}; 
    \draw[fill=gray] (2.449541284,1.5) circle [radius=0.05] node[left] {jl};
    \draw[fill=azure] (3,1.5) circle [radius=0.05] node[right] {\textcolor{azure}{py}}; 
    \draw[fill=azure] (2.481308411,3) circle [radius=0.05] node[above] {\textcolor{azure}{\cpp}};
    \draw[fill=azure] (2.327102804,2) circle [radius=0.05] node[above] {\textcolor{azure}{Fortran}}; 
    \draw[fill=azure] (2.607476636,1.5) circle [radius=0.05] node[below] {\textcolor{azure}{jl}};
    \end{tikzpicture}
    \caption{Heat equation}
    \label{fig:cocomo:heat}
    \end{subfigure}
    \hfill
    \begin{subfigure}[t]{0.28\textwidth}
    \begin{tikzpicture}[scale=0.7, transform shape]
    \draw[help lines, color=gray!30, dashed] (-0.1,-0.1) grid (2.9,2.9);
    \draw[->,thick,cadetgrey] (0,0)--(3,0) node[right]{Difficult};
    \draw[->,thick,cadetgrey] (0,0)--(0,3.1) node[above,cadetgrey]{Good};
    \node[left,cadetgrey] at (0,0) {Easy};
    \node[below,cadetgrey] at (0,0) {Poor};
    \draw[->,thick,cadetgrey] (0,0)--(3,0) node[right]{Difficult};
    \draw[->,thick,cadetgrey] (0,0)--(0,3.1) node[above,cadetgrey]{Good};
    \draw[fill=gray] (2.295454545,3) circle [radius=0.05] node[left] {py}; 
    \draw[fill=gray] (3,3) circle [radius=0.05] node[right] {\cpp};
    \draw[fill=gray] (2.045454545,1.5) circle [radius=0.05] node[left] {jl}; 
    \draw[fill=gray] (2.340909091,3) circle [radius=0.05] node[above] {Fortran};
    \end{tikzpicture}
    \caption{Matrix Multiplication}
    \label{fig:cocomo:matrix}
    \end{subfigure}

    \begin{subfigure}[t]{0.28\textwidth}
    \begin{tikzpicture}[scale=0.7, transform shape]
    \draw[help lines, color=gray!30, dashed] (-0.1,-0.1) grid (2.9,2.9);
    \draw[->,thick,cadetgrey] (0,0)--(3,0) node[right]{Difficult};
    \draw[->,thick,cadetgrey] (0,0)--(0,3.1) node[above,cadetgrey]{Good};
    \node[left,cadetgrey] at (0,0) {Easy};
    \node[below,cadetgrey] at (0,0) {Poor};
    \draw[->,thick,cadetgrey] (0,0)--(3,0) node[right]{Difficult};
    \draw[->,thick,cadetgrey] (0,0)--(0,3.1) node[above,cadetgrey]{Good};
    \draw[fill=gray] (2.805755396,0) circle [radius=0.05] node[below] {py}; 
    \draw[fill=gray] (3,3) circle [radius=0.05] node[above] {\cpp};
    \draw[fill=gray] (2.676258993,1.5) circle [radius=0.05] node[left] {jl}; 
    \draw[fill=gray] (2.978417266,1) circle [radius=0.05] node[above] {Fortran};
    \end{tikzpicture}
    \caption{DGEMM}
    \label{fig:cocomo:dgemm}
    \end{subfigure}
    \hspace{2cm}
    \begin{subfigure}[t]{0.28\textwidth}
    \begin{tikzpicture}[scale=0.7, transform shape]
    \draw[help lines, color=gray!30, dashed] (-0.1,-0.1) grid (2.9,2.9);
    \draw[->,thick,cadetgrey] (0,0)--(3,0) node[right]{Difficult};
    \draw[->,thick,cadetgrey] (0,0)--(0,3.1) node[above,cadetgrey]{Good};
    \node[left,cadetgrey] at (0,0) {Easy};
    \node[below,cadetgrey] at (0,0) {Poor};
    \draw[->,thick,cadetgrey] (0,0)--(3,0) node[right]{Difficult};
    \draw[->,thick,cadetgrey] (0,0)--(0,3.1) node[above,cadetgrey]{Good};
    \draw[fill=gray] (2.771428571,1.5) circle [radius=0.05] node[left] {py}; 
    \draw[fill=gray] (3,2) circle [radius=0.05] node[right] {\cpp};
    \draw[fill=gray] (2.771428571,1.5) circle [radius=0.05] node[below] {jl}; 
    \draw[fill=gray] (2.628571429,2) circle [radius=0.05] node[above] {Fortran};
    \end{tikzpicture}
    \caption{STREAM Triad}
    \label{fig:cocomo:stream}
    \end{subfigure}   
    \caption{Two-dimensional classification using the estimated schedule effort of the COCOMO model (\textbf{easy} vs \textbf{difficult}) and the quality of the code using compilation, execution, and correctness (\textbf{poor} vs \textbf{good}). The blue values show the results from ChatGPT~\cite{diehl2024evaluating} and the black values the results for Deep Seek. The Python and Julia data points are tagged with the common file endings \texttt{py} and \texttt{jl}, respectively.}
    \label{fig:efforts:all}
\end{figure}
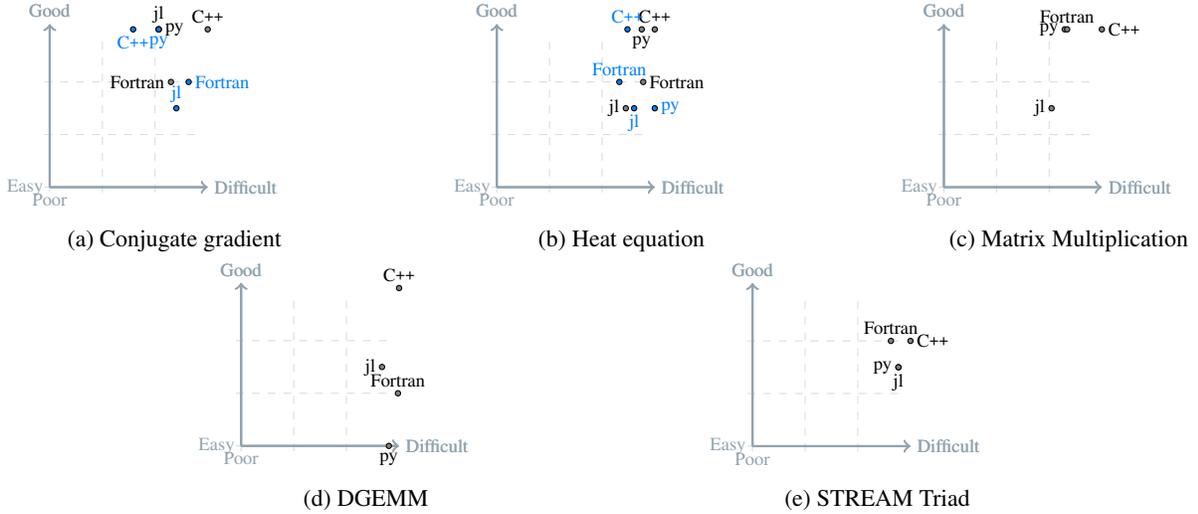

\subsection{Programming efforts }
Similar to the quality, the effort to write the code generated by a programmer is investigated. One way to estimate that effort is by using the \textbf{Co}nstructive \textbf{Co}st \textbf{Mo}del (COCOMO)~\cite{5010193,1237981}. The COCOMO model does not take parallelism into account and ignores features like synchronization. Starting in the 90s, the HPC community discussed the development of a similar model for parallel and distributed applications. No such model has been proposed as of the time of this writing. One attempt was made to add parallel programming to the COCOMO \textit{II} model~\cite{miller2018applicability}. 

We used the tool \textit{scc}\footnote{\url{https://github.com/boyter/scc}} to estimate the human effort to produce the code.
We use Tables~\ref{tab:evaluation:overview},\ref{tab:quality:2} to quantify the quality of the software from \textbf{poor} to \textbf{good}. Figure~\ref{fig:efforts:all} shows the estimated programming efforts and quality of the code. For the conjugate gradient the \cpp\ code needed more effort. For the heat equation the effort was more balanced. For the matrix multiplication the \cpp\ code required the most effort. For DGEMM and STREAM Triad the effort was more balanced. 

\section{Performance evaluation}
\begin{figure}[tb]
    \centering
    \begin{subfigure}[t]{0.45\textwidth}
\begin{tikzpicture}[scale=0.65, transform shape]
\begin{axis}[grid,xlabel=\# cores,ylabel=processed points per second,title=Parallel heat equation (x86-AMD),ymode=log,log basis y={2},xmode=log,
       log basis x={2}]
\addplot[black,mark=square*] table [x expr=\thisrowno{0},y expr={10000000/\thisrowno{1}/1000},col sep=comma] {heat-amd.csv};
\addplot[black,mark=diamond*] table [x expr=\thisrowno{0},y expr={10000000/\thisrowno{1}/1000},col sep=comma] {heat-amd-f.csv};
\addplot[black,mark=*] table [x expr=\thisrowno{0},y expr={10000000/\thisrowno{1}/1000},col sep=comma] {heat-amd-py.csv};
\addplot[black,mark=triangle*] table [x expr=\thisrowno{0},y expr={10000000/\thisrowno{1}/1000},col sep=comma] {heat-amd-julia.csv};
\end{axis}
\end{tikzpicture}
    \caption{}
    \label{fig:performance:heat}
    \end{subfigure}
    \hfill
        \begin{subfigure}[t]{0.45\textwidth}
\begin{tikzpicture}[scale=0.65, transform shape]
\begin{axis}[grid,xlabel=\# cores,ylabel=processed matrix entries per second,title=Parallel Matrix Multiplication (x86-Intel),ymode=log,log basis y={2},xmode=log,
       log basis x={2}]
\addplot[black,mark=square*] table [x expr=\thisrowno{0},y expr={100000000/\thisrowno{1}},col sep=comma] {matrix-riscv-cpp.csv};
\addplot[black,mark=diamond*] table [x expr=\thisrowno{0},y expr={100000000/\thisrowno{1}},col sep=comma] {matrix-riscv-f.csv};
\addplot[black,mark=*] table [x expr=\thisrowno{0},y expr={100000000/\thisrowno{1}},col sep=comma] {matrix-riscv-py.csv};
\addplot[black,mark=triangle*] table [x expr=\thisrowno{0},y expr={100000000/\thisrowno{1}},col sep=comma] {matrix-riscv-jl.csv};
\end{axis}
\end{tikzpicture}
    \caption{}
    \label{fig:performance:matrix}
    \end{subfigure}

    \begin{subfigure}[t]{\textwidth}
    \centering
        \includegraphics[width=0.5\textwidth,trim={0.25em 0 0 0.25em},clip]{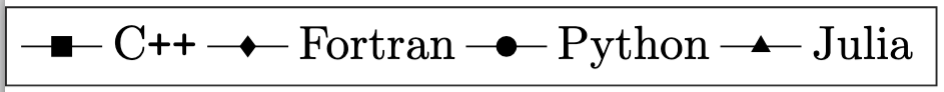}
    \end{subfigure}

\begin{subfigure}[t]{0.45\textwidth}
        \begin{tikzpicture}[scale=0.65, transform shape]
\begin{axis}[
    ybar,
    enlargelimits=0.15,
    legend style={at={(0.5,0.95)},
    anchor=north,legend columns=-1},
    ylabel={GFLOP/s},
    symbolic x coords={512,1024,2048},
    xtick=data,
    ymax=6,
    nodes near coords,
    nodes near coords align={vertical},
    xlabel= row and colnum size,
    title= DGEMM (A64FX)
    ]

\addplot[black,fill=black] coordinates {(512,0.1) (1024,0.2) (2048,0.4) };

\addplot[gray,fill=gray] coordinates {(512,0)(1024,0) (2048,0)  };

\addplot[gray,fill=white] coordinates {(512,0.3) (1024,0.3) (2048,0.3)  };

\addplot[gray,fill=blue] coordinates {(512,2.5)(1024,4.5) (2048,4.1)  };
\legend{\cpp,Fortran,Python,Julia}
\end{axis}
\end{tikzpicture}
    \caption{}
    \label{fig:performance:dgemm}
    \end{subfigure}
    \hfill
        \begin{subfigure}[t]{0.45\textwidth}
        \begin{tikzpicture}[scale=0.65, transform shape]
\begin{axis}[
    ybar,
    enlargelimits=0.15,
    legend style={at={(0.5,0.95)},
    anchor=north,legend columns=-1},
    ylabel={GB/s},
    symbolic x coords={10000,100000,1000000},
    xtick=data,
    ymax=1500,
    nodes near coords,
    nodes near coords align={vertical},
    xlabel=\# elements,
    title= Stream triad (Grace Grace)
    ]

\addplot[black,fill=black] coordinates {(10000,24) (100000,473) (1000000,348) };

\addplot[gray,fill=gray] coordinates {(10000,40) (100000,300) (1000000,1200) };

\addplot[gray,fill=white] coordinates {(10000,0.2) (100000,2.2) (1000000,20) };

\addplot[gray,fill=blue] coordinates {(10000,52) (100000,70) (1000000,69) };
\legend{\cpp,Fortran,Python,Julia}
\end{axis}
\end{tikzpicture}
    \caption{}
    \label{fig:performance:stream}
    \end{subfigure}

    \caption{Performance measurements: (\subref{fig:performance:heat}) parallel heat equation solver on x86-AMD, (\subref{fig:performance:matrix}) parallel matrix multiplication on x86-Intel, (\subref{fig:performance:dgemm}) DGEMM on Arm A64FX, and (\subref{fig:performance:stream}) stream triad on Arm Grace Hopper.}
    \label{fig:enter-label}
\end{figure}
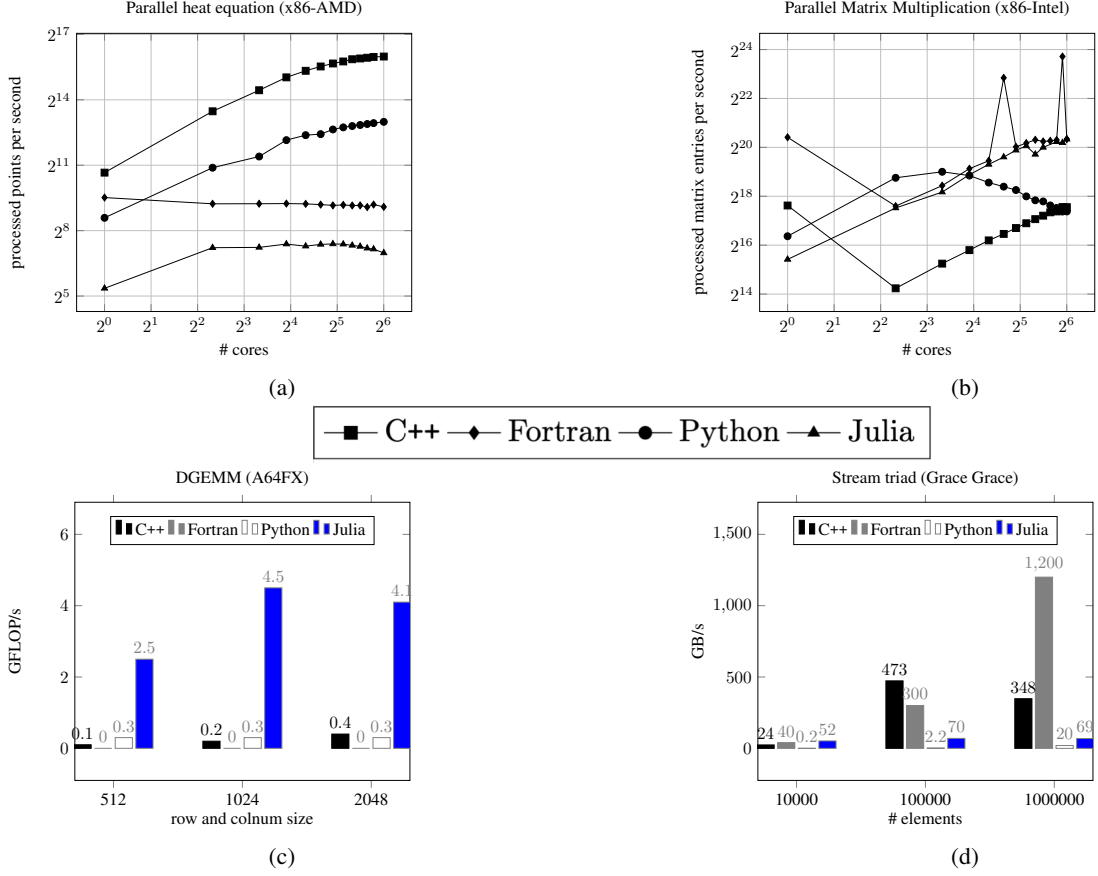

For the performance evaluation, the following versions were used: g\texttt{++}/gfortran 13/14, Julia 1.11.3 and Python 3.11. Recall that we just fixed compilation and runtime errors in the code, but did not improve the parallel implementations. Figure~\ref{fig:performance:heat} shows the scaling for the generated codes on \textit{AMD EPYC 7763} (x86) from one thread to 64 cores. We used $10,000,000$ nodes and changed the length of the domain $L$ to $100,000$. The \cpp\ code and Python code scaled with the number of cores. The Fortran and Julia code showed some speedup from a single core to five cores. After that, the code did not benefit from additional cores. Figure~\ref{fig:performance:matrix} shows the scaling from one core to 64 cores on \textit{Intel Xeon Platinum 8358} (x86). The matrix size was $10,000 \times 10,000$ with $100,000,000$ elements. Here, only the Julia, code scaled. Python, \cpp\, and Fortran showed strange behavior. Figure~\ref{fig:performance:dgemm} shows the DGEMM benchmark for matrices with 512, 1024, and 2048 rows and columns and a block width of 64 on Arm A64FX. These numbers were generated by the LLM. The Fortran code was very slow and the GFLOP/s were in the range of 0.02. The Python code reported around the same values for all sizes. The \cpp\ and Julia code had increasing numbers. However, the Julia codes reported GFLOP/s for a BLAS implementation and these values were hundreds times bigger. Thus, all of the codes had a very poor performance. Figure~\ref{fig:performance:stream} shows the performance for stream triad for $10^4$, $10^5$, and $10^6$ elements on ARM Grace Grace using 72 cores. Here, all code increased performance with the array size. For the Fortran code, we used opencoarrays with openmpi 4.1 as the coarray implementation. Table~\ref{tab:peak:performace} shows the peak performance for the parallel heat equation and matrix multiplication.

\sisetup{scientific-notation = true,round-mode=places,round-precision=1}
\begin{table}[tb]
    \centering
    
    \caption{Peak performance for the parallel heat equation and parallel matrix multiplication as points  processed per second and matrix entries processed per second, respectively. The number in parentheses are the number of cores at the peak performance.}
    \begin{tabular}{l|cccc}\toprule
     Example & \cpp & Fortran & Python & Julia  \\\midrule
    \rowcolor{lightgray} Heat equation & \num{64790338} (64) & \num{604750} (15)  & \num{8127137} (64)  & \num{166899} (15)   \\
     Matrix multiplication & \num{192661} (60) & \num{7502607} (25) & \num{169678} (60) & \num{1225919} (55)     \\\bottomrule
    \end{tabular}
    \label{tab:peak:performace}
\end{table}

\section{Discussion and Conclusion}
The results of this study reveal both the potential and limitations the LLM model DeepSeek in HPC. By evaluating the performance of generated code across different benchmarks, we identified key insights regarding code quality, scalability, and parallel execution.

One important observation is that LLMs struggle to write scalable codes. In first two examples, both the \cpp\ and Python codes showed consistent scaling with the increase in the number of cores, while the Fortran and Julia implementations showed limited scalability. This is consistent with prior research on HPC applications, which highlights the challenges of parallelism in certain programming languages. In example three, both Python and Fortran did not scale.
The Stream Triad clearly improved in performance with increasing array sizes for the all generated codes. 
Using COCOMO analysis, the \cpp\  code required more effort in the conjugate gradient and matrix multiplication examples. While the effort was more balanced between all the languages for the heat equation solver and STREAM Triad tasks.

DeepSeek shows promise in generating code across several HPC benchmarks, however the performance of the generated code needs more improvement for HPC tasks. The results also highlight the difficulties in optimizing for parallel execution and memory efficiency when generating code with LLMs. Future work should focus on improving the parallelization and performance of the generated code for HPC tasks like DGEMM and Stream Triad. To conclude, when used alongside traditional programming methods, LLMs can significantly reduce the effort required for code generation, but there is still work to be done to reach the performance levels expected for high-performance applications.

In a future work we will study the performance for distributed applications, like MPI, acceleration cards, and abstraction layers (like Kokkos or SYCL).

\subsection*{Supplementary materials}
The generated source code is available on GitHub\footnote{\url{https://github.com/NoujoudNader/AiCode\_DeepSeek}} or Zenodo\footnote{\url{https://doi.org/10.5281/zenodo.14968599}}, respectively.

\subsection*{Acknowledgments}
{\footnotesize This work was supported by the U.S. Department of Energy through the Los Alamos National Laboratory. Los Alamos National Laboratory is operated by Triad National Security, LLC, for the National Nuclear Security Administration of U.S. Department of Energy (Contract No. 12345) LA-UR-25-22174}
%
%
%
\bibliographystyle{splncs04}
\bibliography{references}
\end{document}